\documentclass[twocolumn,numberedappendix,appendixfloats]{emulateapj}
\usepackage{natbib}
\usepackage{graphicx}
\usepackage{dcolumn}
\usepackage{bm}
\usepackage{amsmath}
\usepackage{amssymb}
\usepackage{hyperref}
\usepackage[all]{hypcap}
\usepackage{xcolor}
\usepackage{tabulary}
\shorttitle{On the conservation of the vertical action}
\shortauthors{Vera-Ciro \& D'Onghia}
\graphicspath{{figs/}}

\newcommand{\diskone}{\texttt{disk 1}}
\newcommand{\disktwo}{\texttt{disk 2}}

\hypersetup{
    bookmarks=true,              
    unicode=false,               
    pdftoolbar=true,             
    pdfmenubar=true,             
    pdffitwindow=false,          
    pdfstartview={FitH},         
    pdftitle={Vertical Action},  
    pdfauthor={Vera-Ciro et al.},
    pdfnewwindow=true,           
    colorlinks=true,             
    linkcolor=blue,              
    citecolor=blue,              
    filecolor=blue,              
    urlcolor=blue                
  }

\begin{document}

\title{On the conservation of the vertical action in galactic disks}

\author{Carlos Vera-Ciro\altaffilmark{1}, Elena D'Onghia\altaffilmark{1,2}}

\begin{abstract}
  We employ high-resolution $N$-body simulations of isolated spiral
  galaxy models, from low-amplitude, multi-armed galaxies to Milky
  Way-like disks, to estimate the vertical action of ensembles of stars
  in an axisymmetrical potential.  In the multi-armed galaxy the
  low-amplitude arms represent tiny perturbations of the potential,
  hence the vertical action for a set of stars is conserved, although
  after several orbital periods of revolution the conservation
  degrades significantly.  For a Milky Way-like galaxy with vigorous
  spiral activity and the formation of a bar, our results show that
  the potential is far from steady, implying that the action is not a
  constant of motion. Furthermore, because of the presence of
  high-amplitude arms and the bar, considerable in-plane and vertical
  heating occurs that forces stars to deviate from near-circular
  orbits, reducing the degree at which the actions are conserved for
  individual stars, in agreement with previous results, but also for
  ensembles of stars.  If confirmed, this result has several
  implications, including the assertion that the thick disk of our
  Galaxy forms by radial migration of stars, under the assumption of
  the conservation of the action describing the vertical motion of
  stars.

\end{abstract}


\altaffiltext{1}{Department of Astronomy, University of Wisconsin, 2535
  Sterling Hall, 475 N. Charter Street, Madison, WI 53076, USA.
  \href{mailto:ciro@astro.wisc.edu}{e-mail:ciro@astro.wisc.edu}}

\altaffiltext{2}{Alfred P. Sloan Fellow}

\section{Introduction}
\label{sec:intro}

A stellar system can be fully modeled by a phase-space density
distribution function (DF), whose evolution is described by solutions
of the Boltzmann equation \citep{Binney2008}. Is also a well-known
result that for steady-state potentials, the DF describing the galaxy
can be represented as a function of the integrals of motion of the
system \citep{Jeans1919}. There are several choices one can make to
use then as coordinates of the DF, e.g. energy and angular momentum
\citep{Eddington1915}. Integrated orbits in axisymmetric potentials
indicate the existence of a non-classical isolating integral of motion
\citep{Ollongren1962}, which is closely related to the actions of the
system \citep{Binney1984}.

Besides being a natural choice to describe the evolution in the
phase-space of collisionless stellar systems, the actions are also
adiabatic invariants. This is a
very appealing property of spiral galaxies, where the high frequency
of the vertical motion, compared to in-plane evolution, makes the
action describing the $z$ motion a prime candidate for being a
conserved quantity, even in the presence of spiral arms. This
assumption has been tested by \cite{Solway2012}, who showed that for
isolated galaxies the vertical action is not a constant of motion of
individual stars, and is only conserved in average for samples of
stars, with an intrinsic dispersion of $\sim 20\%$.

It is known that recurring spiral arm-activity in a galactic disk
causes stars to diffuse radially over time in a manner that largely
preserves the overall structure of the disk. \citet{Sellwood2002} used
bi-dimensional stellar disks to show that stars near the corotation
radius of a transient spiral pattern change angular momentum and move
to new radii inwards or outwards with negligible increase of random
motion, causing a process termed radial migration. Following studies
confirmed that this process occurs in 3D simulations of galactic disks
either in isolation or in a cosmological context \citep{Roskar2008,
  Minchev2011, Grand2012, Loebman2011, Kubryk2014, Halle2015,
  Grand2015}.

If radial mixing in galactic disks occurs, this process has
interesting implications for the spread in the age-metallicity
relation observed in the solar neighborhood \citep{Edvardsson1993}. If
stellar migration never occurred, the stars in the solar vicinity
would have metallicities ranging from zero to the present-day value
\citep{Serenelli2009} with a significant correlation between the age
of the stars and their metallicities. However this correlation has
never been found in the current data, raising the possibility that the
stars migrated from their birth radii \citep{Wielen1996}.  Because the
radii of stars at the birth depends on their specific angular
momentum, stellar migration occurs if there is a change of the angular
momentum of the stars.  Changes in the angular momentum for individual
stars can be induced by scattering with giant molecular clouds
\citep{Spitzer1953, Mihalas1981} or transient spiral structures
\citep{Jenkins1990, Jenkins1992}, interference of coherent long-lived
modes and density waves with a bar \citep{Roskar2012}, or by
interactions with satellite galaxies \citep{Kazantzidis2008,
  Villalobos2008, Bird2012}.

In this paper we aim to study in detail the effect of migration on
isolated disks, by using $N$-body simulations for disk galaxies
evolving in isolation, to achieve high-resolution; with the aim of
better determining the contribution of the spiral waves to the heating
of the stellar disk, to the radial migration of stars in the disk, and
ultimately to the formation of a thick disk.

This manuscript is laid out as follows. In
Section~\ref{sec:preliminaries} we introduce the $N$-body simulations
used in our work and describe the method to measure the vertical
action. In Section~\ref{sec:conservation} we discuss to what extent
this quantity is conserved during the evolution of the disk, and in
Section~\ref{sec:thick} we show how this can be used to study the
formation of a thick disk. Finally in Section~\ref{sec:conclusions} we
draw our conclusions.

\section{Numerical Preliminaries}
\label{sec:preliminaries}

\subsection{Simulations}

For this study we employed $N$-body simulations of two galaxy models
labeled as \diskone{} and \disktwo. Each galaxy consisted of a static
dark matter halo and a live rotationally supported stellar disk of
$5\times 10^6$ particles. Because the spiral morphology in stellar
disks depends in detail on the mass distribution of the galaxy and on
the self-gravity of the disk, we selected the structural parameters of
the galaxies such that one galaxy developed a low-amplitude
multi-armed disk (\diskone), whereas the other developed a spiral
morphology more similar to a Milky Way (MW)-like galaxy (\disktwo).
This was achieved by choosing models with very different disk mass
fractions within $2.2$ scale-lengths: $M_{\rm disk}/M_{\rm total}=
0.271$ for \diskone{} and $0.478$ for \disktwo, respectively. By
varying the mass distribution of the disk within $2.2$ scale-lengths,
these models were designed to change the critical length scale
parameter $\lambda_{\rm crit} = 4\pi^2 G \Sigma/\kappa^2$, where
$\Sigma$ is the disk surface mass density and $\kappa$ the epicycle
frequency, and lead to a different spiral morphology
\citep{Toomre1981, Sellwood1984, Carlberg1985, DOnghia2015}.

The parameters describing each galaxy component were assumed to be
independent and both models were constructed similarly to the approach
described in \citep{Hernquist1993, Springel2005}.  The dark matter
mass distribution was modeled assuming the Hernquist profile
\citep{Hernquist1990}:

\begin{equation}\label{eq:dens-halo}
  \rho_{\rm halo}(r)=\frac{M_{\rm halo}}{2\pi R_{\rm halo}^3}\frac{1}{(r/R_{\rm
      halo})(1 + r/R_{\rm halo})^3},
\end{equation}

\noindent with $M_{\rm halo}$ being the total halo mass and $R_{\rm
  halo}$ the scale radius. The effect of the dark halo is modeled
assuming a static potential, this is of course a limitation of our
model since the system never completely relaxes, however, the
complication of adding an alive component comes with the price of
including further scattering to the particles of the disk.  The
stellar disk is modeled in the initial conditions as a thin
exponential surface density profile of scale-length $R_{\rm disk}$ and
total mass $M_{\rm disk}$. The vertical mass distribution of the stars
in the disk is specified by the profile of an isothermal sheet with a
radially constant scale height $z_{\rm disk}$:

\begin{equation}\label{eq:dens-disk}
  \rho_{\rm disk}(R,z) = \frac{M_{\rm disk}}{4\pi R^2_{\rm disk}z_{\rm disk}} {\rm
    e}^{-R/R_{\rm disk}} {\rm sech}^2 \frac{z}{z_{\rm disk}}.
\end{equation}

Model \disktwo{} included a static potential for the bulge described
by the Hernquist model with mass $M_{\rm bulge}$ and scale-length
$R_{\rm bulge}$. We aim to build this galaxy model with a flat
rotation curve consistent with the observations of the MW within two
optical radii ($R\lesssim 14$ kpc) \citep{Reid2014, Bovy2009}. We
noted that this requirement is satisfied by adopting a Hernquist
profile for the dark halo with a large scale radius, and a large total
mass. However, the dynamical range of interest of this study is a few
optical radii, thus the mass distribution outside this radius does not
affect our analysis.  Table \ref{tab:params} summarizes the structural
parameters adopted for each galaxy model in this study.

\begin{table}
  \begin{center}
    \caption{Structural properties of the galaxy models}
    \label{tab:params}
    \begin{tabular}{l|c|c} \hline
      & \diskone & \disktwo \\ \hline
      $M_{\rm disk}$ $(10^{10}\; {\rm M}_{\odot})$ & 1.905 & 4.000 \\
      $R_{\rm disk}$ (kpc) & 3.130 & 2.500 \\
      $z_{\rm disk}/R_{\rm disk}$ & 0.100 & 0.150 \\
      $M_{\rm halo}$ $(10^{12}\; {\rm M}_{\odot})$ & 0.933 & 9.500 \\
      $R_{\rm halo}$ (kpc) & 29.775 & 130.0 \\
      $M_{\rm disk}/M_{\rm total}(2.2 R_{\rm disk})$ & 0.271 & 0.479 \\
      $M_{\rm bulge}$  $(10^{10}\; {\rm M}_{\odot})$ & \ldots & 1.400 \\
      $R_{\rm bulge}$ (kpc) & \ldots & 0.350 \\ \hline \hline
    \end{tabular}
  \end{center}
\end{table}

\begin{figure}
  \begin{center}
    \includegraphics[width=0.49\textwidth]{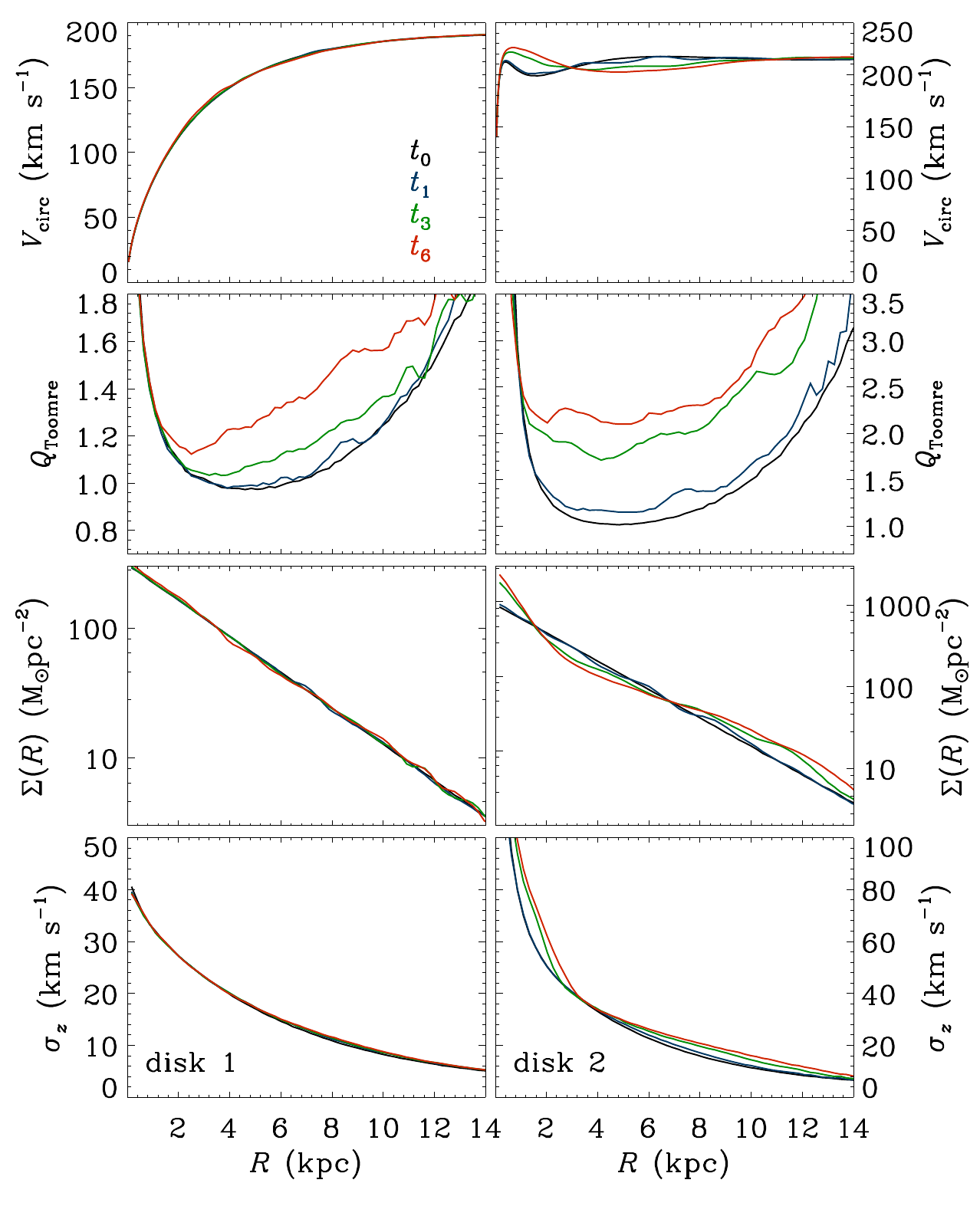}
  \end{center}
  \caption{Radial profiles of total circular velocity, $V_{\rm cir}$,
    Toomre parameter, $Q_{\rm Toomre}$, surface mass density,
    $\Sigma$, and vertical velocity dispersion, $\sigma_z$, for the
    two studied galaxy models: \diskone{} (left) and \disktwo{}
    (right).  $t_n$ refers to the time of integration in our
    simulations in Gyr, with $t_0$ being the the initial time and
    $t_6$ corresponding to $6$ Gyr.}
  \label{fig:vcirc}
\end{figure}

Fig.~\ref{fig:vcirc} shows the azimuthally averaged radial profiles
for the total circular velocity $V_{\rm circ}$ (stars and dark
matter), the Toomre parameter $Q_{\rm Toomre}$ of the stellar disk,
the disk surface density $\Sigma$ and its vertical velocity dispersion
$\sigma_z$. The profiles are displayed at four different times,
spanning 6 Gyr of evolution. In what follows we use the notation $t_n$
to refer to the time of integration in our simulations in Gyr, with
$t_0$ thus labeling the the initial time, and $t_6$ corresponding to
$6$ Gyr.

We note that in the multi-armed galaxy (\diskone{} model), $Q_{\rm
  Toomre}$ increases with time, as a consequence of the increase of the
radial velocity dispersion. This is expected, since in-plane heating
occurs as a consequence of the spiral structure activity, while the
vertical structure remains unchanged.

This is, however, not the case for our \disktwo{} run. A bar
instability grows at the center of the galaxy after $\sim 3$ Gyr.
Despite the flat rotation curve, the surface mass density of the disk
changes considerably with time and the vertical velocity dispersion
increases by $\sim 15\%$ in the shown radial range.

\begin{figure}
  \begin{center}
    \includegraphics[width=0.49\textwidth]{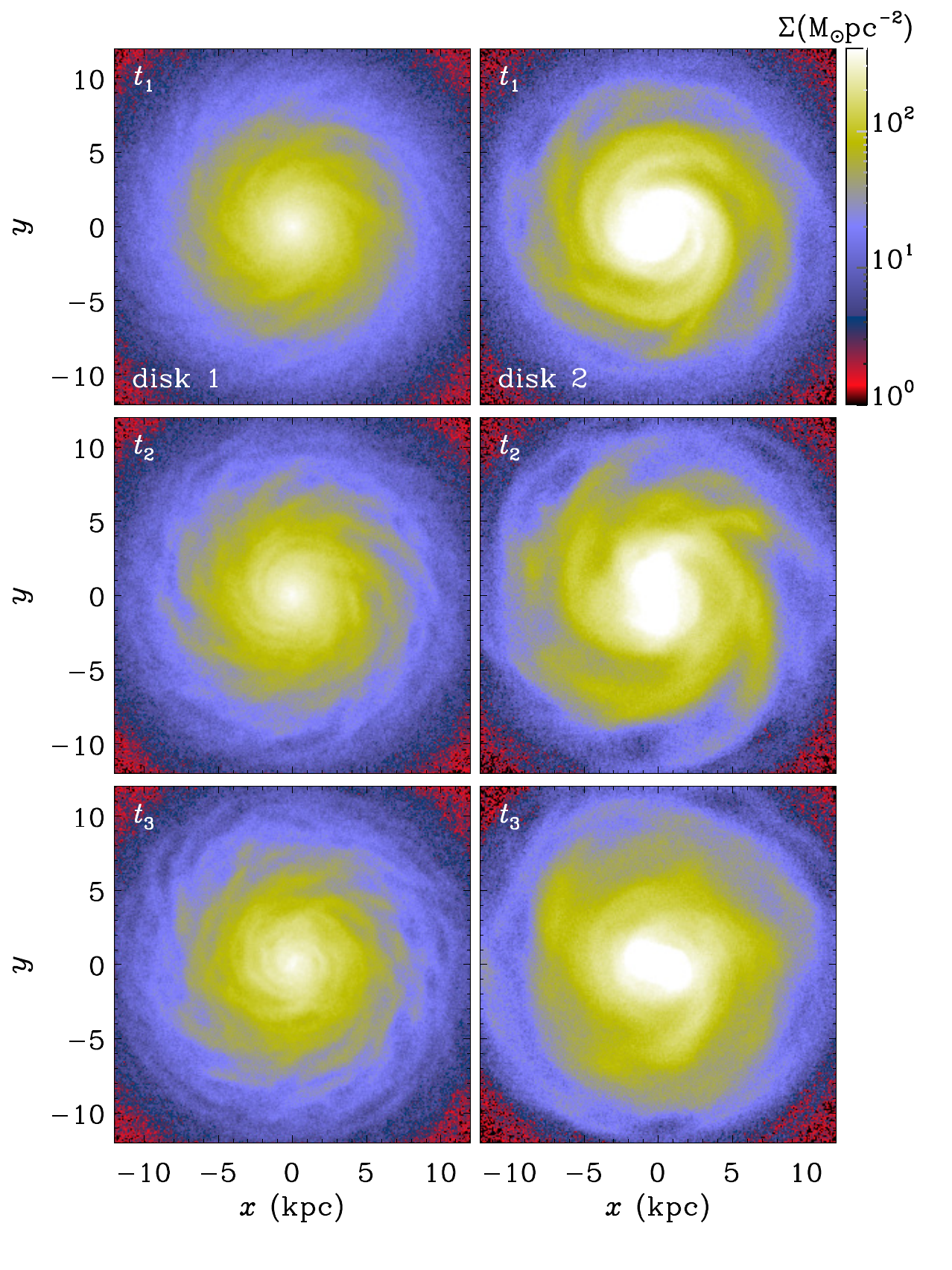}
  \end{center}
  \caption{Time sequence of density projections of the face-on views
    of stellar disks presented in this study. The \diskone{}
    simulation develops a large number of low-amplitude spiral
    structure that perpetuate for at least 6 Gyr (left panels);
    \disktwo{} develops approximately a four-fold rotational symmetry
    and becomes bar-unstable after $\sim 3$ Gyr.}
\label{fig:faceon}
\end{figure}

In Fig.~\ref{fig:faceon} a sequence in time of the face-on views of
the stellar disk of both models of galaxies are displayed. The colored
bar indicates the values for the surface density of the disk. As
expected in the context of swing amplification theory, the two models
show very different spiral morphologies.  Our \diskone{} model grows a
large number of low-amplitude spiral arms that self-perpetuate for at
least 6 Gyr \citep{DOnghia2013}. After 1 Gyr \disktwo{} already
developed four high-amplitude spiral arms at two scale-lengths ($\sim 5$
kpc), in remarkable agreement with the number of arms predicted for a
disk with 50\% disk fraction \citep{DOnghia2015, Pettitt2015}. Later
in time this disk model develops a bar \citep{Merritt1994,
  Martinez2006, Athanassoula2008, Yurin2015}, which becomes the
dominant non-axisymmetric feature after $\sim 4$ Gyr of evolution.

Note that the efficiency of the stellar radial migration by recurrent
spiral activity in the multi-armed galaxy (\diskone{} model) has been
recently studied by \cite{VeraCiro2014}. As first shown in
\cite{Sellwood2002}, the authors confirmed that spiral activity in a
low-mass disc causes stars to diffuse radially, with stars near
corotation of a spiral pattern exchanging angular momentum and moving to a
new radius without adding random motion \citep[see
also][]{Daniel2015}. Indeed, the mechanism of scattering of stars at
corotation predicts a lack of significant disk heating and leaves the
surface density profile unchanged. These two features are reproduced
in our \diskone{} model as shown in Fig.~\ref{fig:faceon} (left panels).
In the MW-like galaxy model (\disktwo) the bar formation causes
greater angular momentum changes than those of a multi-armed spiral
structure \citep{Minchev2010}. Furthermore, because the bar persists
after its formation, the associated angular momentum changes might
differ from those caused by recursive spiral arms. Therefore the
surface density profile of the disk is expected to change. Some disk
heating is also expected to occur with time.

\subsection{Measuring the vertical action}

The action-angle variables can be used to describe the evolution of
orbits in static potentials.  The actions, $j$, are constants of motion
in the unperturbed field, or when the potential changes slowly so that
they are adiabatic invariants \citep{Arnold1978,
  Goldstein2002}.  The vertical action is defined as:

\begin{equation}\label{eq:action}
  j_z \equiv \frac{1}{2\pi}\int {\rm d}z\; {\rm d}v_z = \frac{1}{2\pi}\oint {\rm d}z\; v_z,
\end{equation}

\noindent where $z$ and $v_z$ describe the position and velocity along
the orbit, respectively.  There are a number of ways to estimate of
the vertical action of a star particle in a nearly axisymmetric
potential. \cite{Solway2012}, for instance, calculated the quadrature
in Eq.~\eqref{eq:action} by measuring the area enclosed by the orbit
in the $z-v_z$ plane. We adopted here the prescription as described in
Appendix \ref{sec:appendix}. This method allows us to identify and
reject stars trapped in resonances, a case where the adiabatic
invariance breaks \citep{Pfenniger1984, Pfenniger1991}.

\begin{figure}
  \begin{center}
    \includegraphics[width=0.49\textwidth]{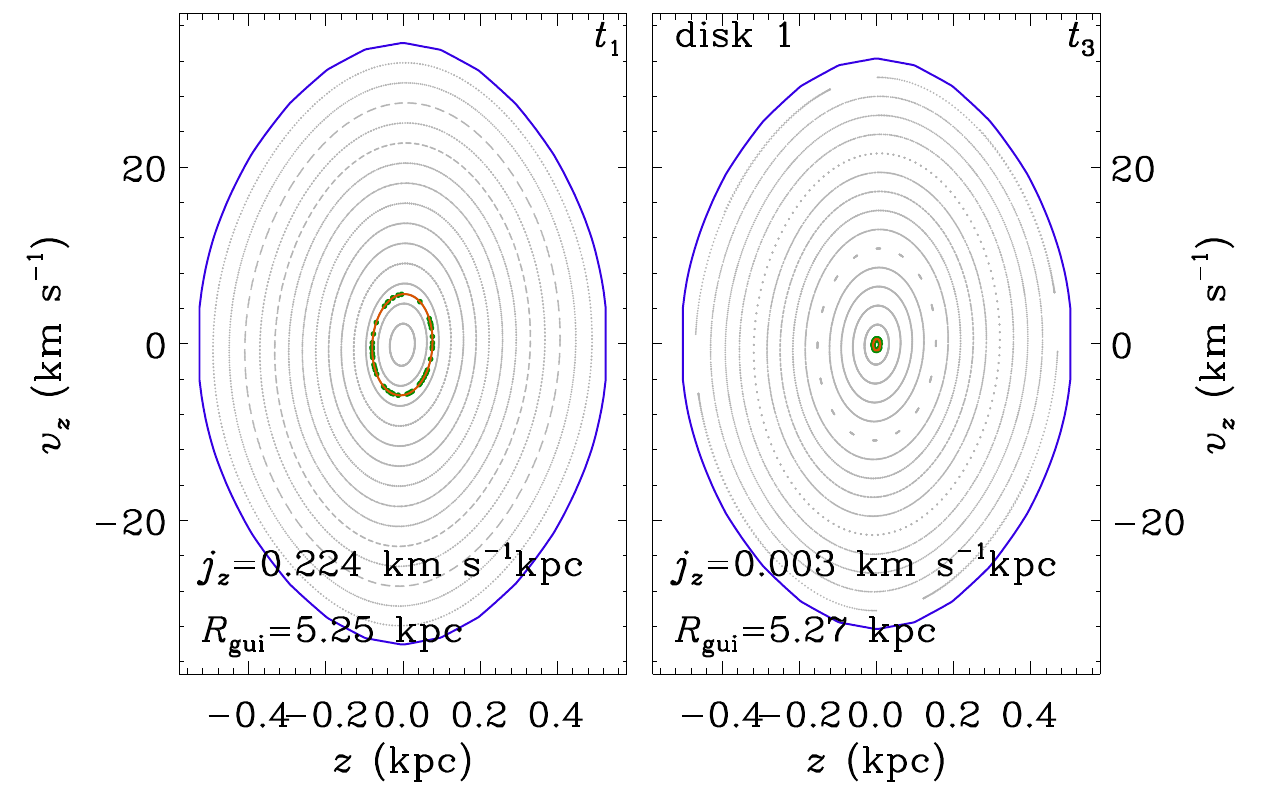}
  \end{center}
  \caption{Surface of section for a star particle measured at two different
    times. In the left panel the simulation has already developed
    spiral arms, the right panel is the surface of section after 2
    Gyr. In this specific example, $R_{\rm gui}(t_1) \approx R_{\rm gui}(t_3)$
    indicates that the particle has not migrated significantly. 
    However, its vertical action has changed by a factor $\sim80$
    ($\delta j_z = 4.4$). For each case the solid red line displays the
    closed path used to define the action and the blue line indicates the
    zero-velocity curve.}
  \label{fig:sos}
\end{figure}

In Fig.~\ref{fig:sos} we show an example for a random particle in our
\diskone{} simulation at two different times. The green points show the
surface of section of the integrated orbit and the red line is the
connected loop used to calculate Eq.~\eqref{eq:action}. The blue line
represents the zero-velocity curve, and gray points are other orbits,
shown here to depict the structure of the phase-space in our runs. The
area defined by the red-loop in Fig.~\ref{fig:sos} is the vertical
action $j_z$, for this particular example, the vertical action
measured at $t_1$ is a factor of $\sim 80$ larger than it is at time
$t_3$, this is clear evidence that the value of $j_z$ is not
conserved for individual orbits for slow-varying, low-amplitude
perturbations, such as the ones developed in our \diskone{} run.

\section{A conserved quantity?}
\label{sec:conservation}

\begin{figure}
  \begin{center}
    \includegraphics[width=0.49\textwidth]{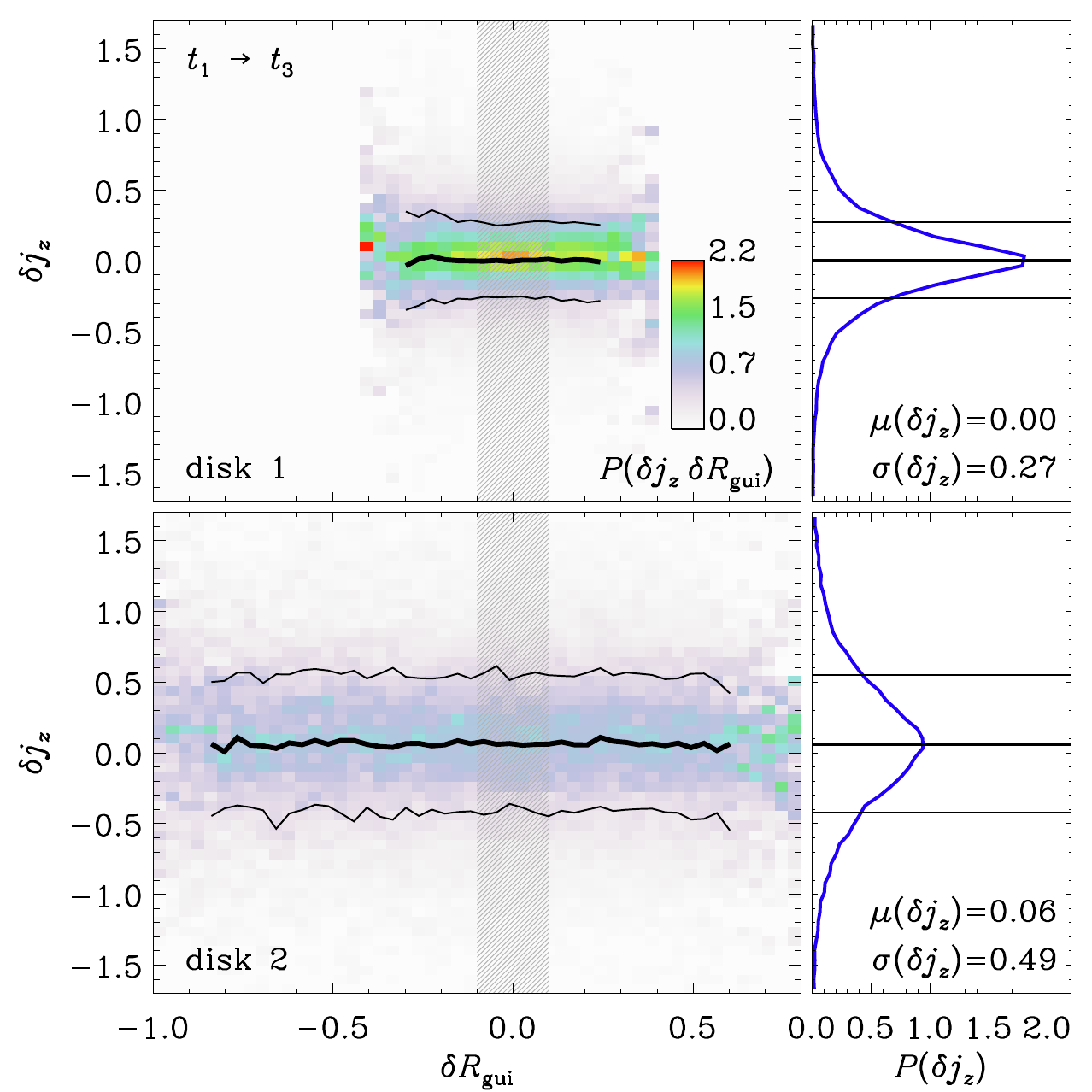}
  \end{center}
  \caption{Probability distribution of fractional changes of the
    vertical action $j_z$, as a function of changes in guiding radius
    $R_{\rm gui}$ for particles selected at 1 Gyr with $R_{\rm gui} =
    5.00 \pm 0.25$ kpc. The distribution of fractional changes in
    $j_z$ does not depend on how efficiently particles migrate. The
    median and $1\sigma$ equivalent dispersion are represented with
    solid lines in the left panels. The right panels show the
    marginalized distribution of changes in the vertical action.}
  \label{fig:djz-drgui}
\end{figure}

Our previous calculations demonstrated that the vertical action $j_z$
is not a conserved quantity for individual stars. This result agrees
with findings of previous studies \citep{Solway2012}. We investigated
this property in our \diskone{} and \disktwo{} simulations.  After 1
Gyr of evolution an ensemble of stars with guiding radius: $R_{\rm
  gui} = 5.00\pm 0.25$ kpc was selected and the relative changes in
the vertical action were computed at a later time (See
Appendix~\ref{sec:appendix}). This radius is close to two disk
scale-lengths in both galaxy models. The vertical action changed
significantly over a period of time as shown in
Fig.~\ref{fig:sos}. Thus, its change from time $t$ to a later time $t'
> t$ was quantified in the following way:

\begin{equation}
\delta j_z \equiv \ln \frac{j_z(t')}{j_z(t)}.
\end{equation}

In addition, we noted that over the same period of time a significant
radial migration occurs, with a consequent change of the guiding
radii. Thus, a similar definition was also applied to $\delta R_{\rm
  gui}$.  Fig.~\ref{fig:djz-drgui} (top panel) displays the outcome of
our study for the multi-armed galaxy \diskone{} over a period of 2
Gyr. In agreement with previous studies, our findings suggest the
vertical action is a conserved quantity on average for an ensemble of
stars. Note that the modest fractional changes of $\delta j_z$ in our
simulation match the values reported by
\citet{Solway2012}\footnote{$\ln x \approx x - 1$, for $x \sim 1$.}
Furthermore, Fig.~\ref{fig:djz-drgui} indicates that any variation in
the vertical action, $\delta j_z$, is independent of the radial
migration, measured by changes in guiding radii, $\delta R_{\rm gui}$.
Next, we computed the median of the fractional change in the vertical
action $\mu (\delta j_z)$ and the $1\sigma$ equivalent dispersion
around the median $\sigma (\delta j_z)$.  We obtained $\mu (\delta j_z)=
0$, with standard deviation $\sigma (\delta j_z)=0.28$ (displayed as
the solid black line in Fig.~\ref{fig:djz-drgui}), also in agreement
with the values reported in \cite{Solway2012}.

The same analysis performed on the more massive \disktwo{} produces
different results as shown in Fig.~\ref{fig:djz-drgui} (bottom panel).
The median of fractional changes in the vertical action is $\mu
(\delta j_z)=0.09$ with a dispersion $\sigma (\delta j_z)=0.64$,
significantly larger than the case of the multi-armed galaxy. Thus,
changes of angular momentum occurred with the formation of the bar,
with significant disk heating, combined with the strong variation
of the potential of the disk, do not conserve the vertical action at
the same level of accuracy.

\begin{figure}
  \begin{center}
    \includegraphics[width=0.49\textwidth]{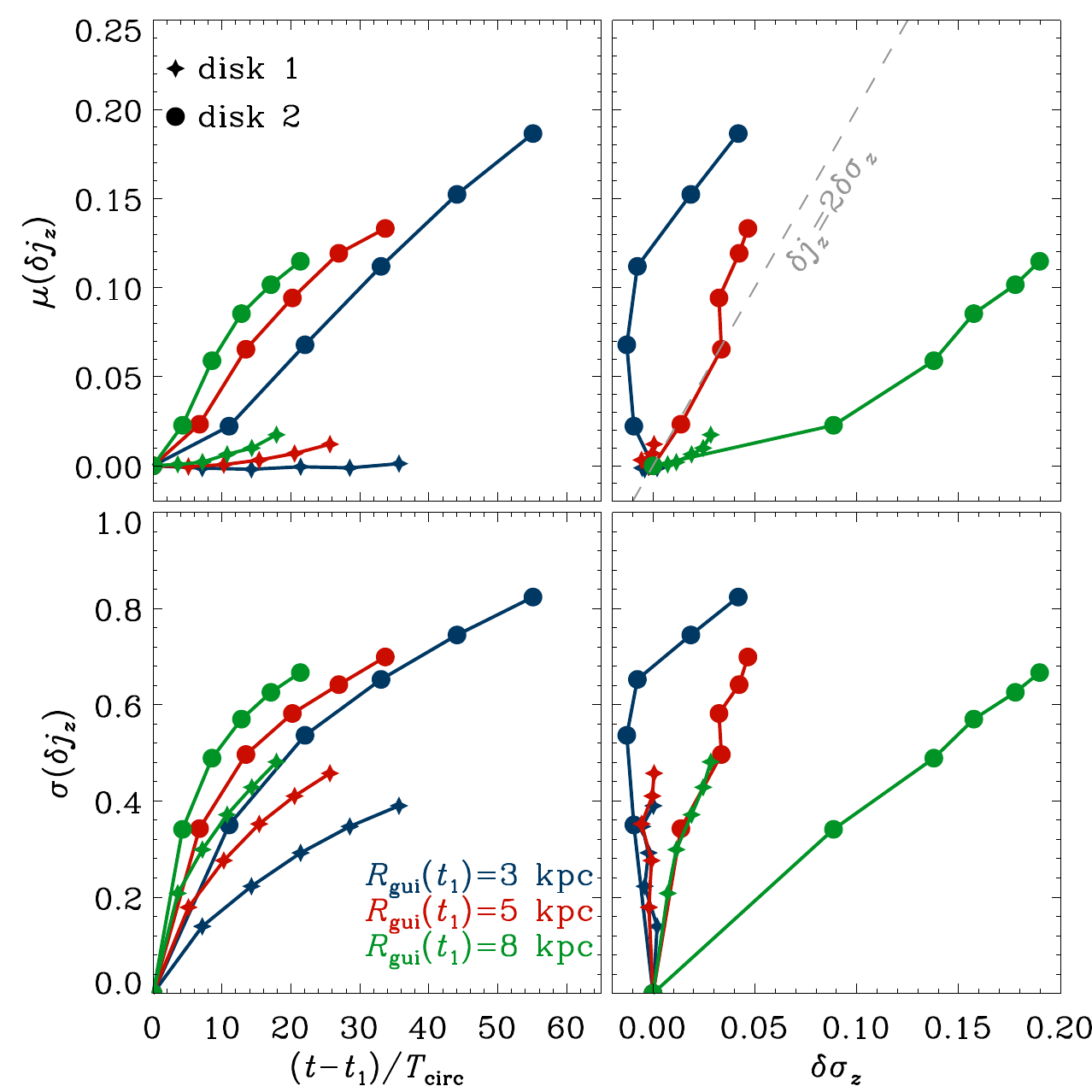}
  \end{center}
  \caption{Left panels: evolution of average changes in the vertical
    action (top panels) and its $1\sigma$ equivalent dispersion
    (bottom panels) as a function of time.  Right panels: the same
    quantities as in the left panels are plotted against the changes
    in vertical velocity dispersion for ensembles of stars selected at
    different radii.}
  \label{fig:djz-t}
\end{figure}

To demonstrate this unexpected outcome, the evolution by time of the
median $\mu(\delta j_z)$ and its dispersion $\sigma (\delta j_z)$ of
fractional changes in the vertical action are displayed in
Fig. ~\ref{fig:djz-t} for the two galaxy models.  Ensembles of stars
are selected at after 1 Gyr of evolution ($t = t_1$) with guiding
radii $R_{\rm gui}=3,5,8$ kpc and followed over the evolution time of
the disk.  While in the multi-armed disk, even after 40 orbital periods
the average variation in the vertical action is never larger than 1\%,
in the MW-like disk galaxy the median reaches values of 15-20\% after
20 orbital periods.

Note that in Fig. ~\ref{fig:djz-t} the dispersion around the median
steadily increases as a function of time for both models (left panels),
suggesting that the vertical action of an ensemble of stars tends to
be less conserved as time progresses. This results holds for both the
multi-armed disk and the MW-like disk.

We also checked whether the vertical heating of the disk, measured by the
vertical velocity dispersion of the ensemble of stars, could produce
changes in the average vertical action. Fig.~\ref{fig:djz-t}
plots the median $\mu(\delta j_z)$ and dispersion $\sigma (\delta
j_z)$ of fractional changes in the vertical action against the
corresponding variation in the velocity dispersion of the vertical
component.

To test for the origin of the changes in the average action we note
that the vertical energy of samples of stars proportionally depends on
the vertical velocity dispersion $\langle j_z \rangle \sim \sigma_z^2$
(cf. Eq.~\eqref{eq:hdisk}), and so if the average vertical action
increases as a consequence of the vertical heating in the disk, then
$\delta j_z = 2\delta \sigma_z$. The multiplying factor in this
approximation is the vertical epicycle frequency, which remains mostly
constant in time (See Fig.~\ref{fig:vcirc}). Fig.~\ref{fig:djz-t}
shows that this is not the case, and that the increase in dispersion
of the changes in the vertical action is not merely determined by the
vertical heating of the disk.

\section{Implications for the vertical structure}
\label{sec:thick}

\begin{figure}
  \begin{center}
    \includegraphics[width=0.49\textwidth]{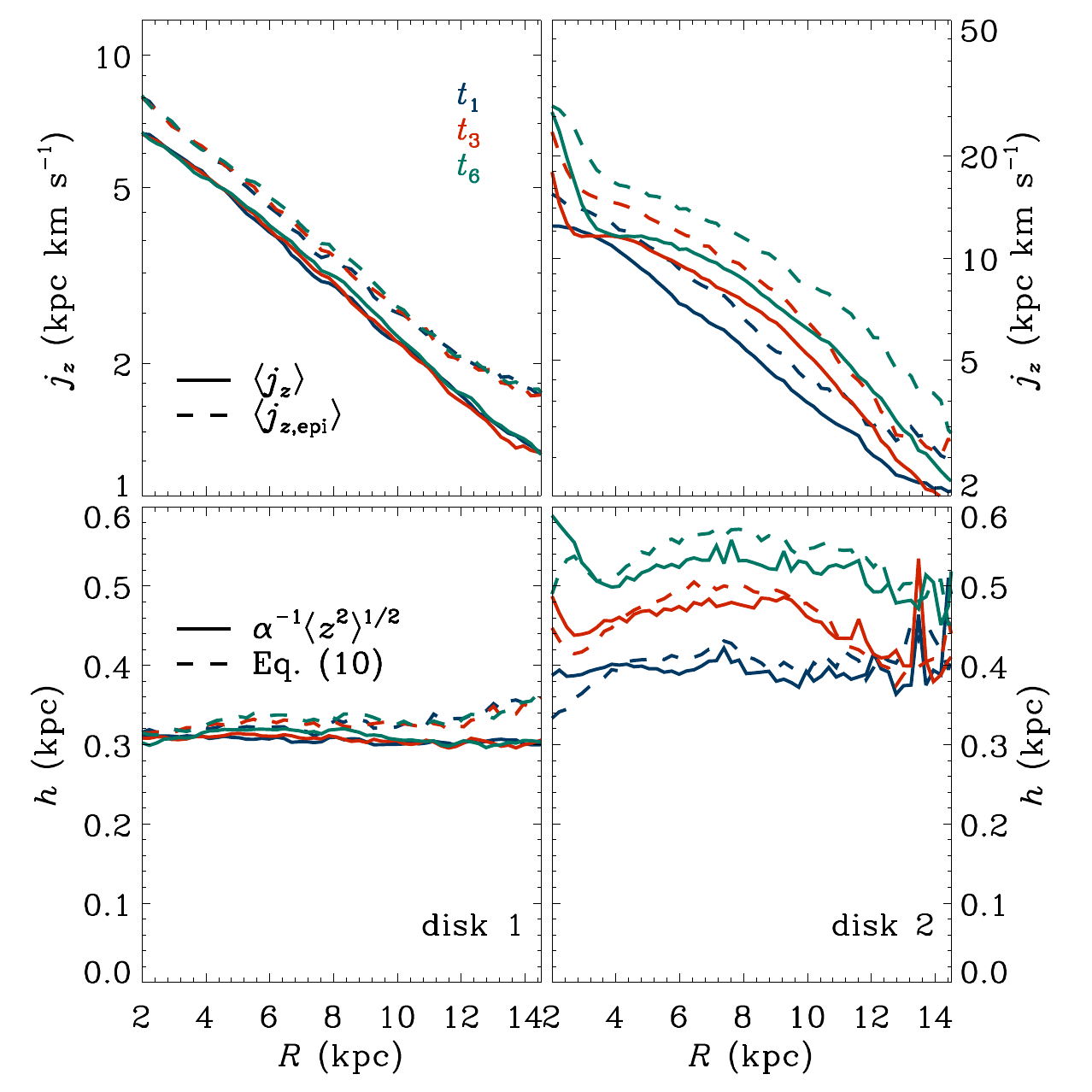}
  \end{center}
  \caption{Top: average vertical action as a function of radius, the
    solid lines are the values calculated using the method described
    in the previous section (Eq.~\eqref{eq:action}) whereas the dashed
    lines correspond to the epicycle approximation
    (Eq.~\eqref{eq:energyz}). The difference between these two values
    is $\sim 15\%$. Bottom: thickness of the disk calculated as the
    RMS values of the $z$-coordinates (solid), and by solving
    Eq.~\eqref{eq:hdisk}.}
  \label{fig:hdisk}
\end{figure}

What does the conservation of the vertical action imply for the
structure and evolution of the stellar disk? It has been argued that
under the assumption of the conservation of the vertical action $j_z$
stars migrating outwards in the disk tend to move far from the
mid-plane and populate the thick disk \citep{Roskar2008,
  Schonrich2009b}. We address this question here.

The epicycle approximation regulates the motion of stars on
near-circular orbits.  In the approximation the vertical motion
of a star is also decoupled from the in-plane motion. The motion
perpendicular to the plane is then described by the Hamiltonian
\citep{Binney2008}:

\begin{equation}
  H_z = \frac{1}{2}v_z^2 + \frac{1}{2}\nu^2z^2.
\end{equation}

For an ensemble of star particles at radius $R$, the average vertical
energy is

\begin{equation}\label{eq:energyz}
  \langle H_z\rangle = \frac{1}{2}\sigma_z^2 + \frac{1}{2}\nu^2\langle z^2\rangle,
\end{equation}

\noindent where $\sigma_z$ is the vertical velocity dispersion of the
ensemble.  Note that $\sigma_z$ in Eq.~\eqref{eq:energyz} does not
necessarily refer to the value of velocity dispersion of all stars with
home radii at $R$. Indeed, Eq.~\eqref{eq:energyz} holds also for any
population of stars whose guiding radii have changed over time
(e.g. stars migrating from inner or outer locations within the disk)
and that are used to calculate it. Thus, in the epicycle
approximation, the vertical action is the following $j_z = H_z / \nu$
\citep{Binney2008}:

\begin{equation}\label{eq:hdisk}
  \langle j_{z, \rm epi}\rangle  = \frac{\langle H_z\rangle}{\nu} = \frac{1}{2\nu}(\sigma_z^2 +
  \nu^2\alpha^2 h^2), 
\end{equation}

\noindent where, $\alpha^2 h^2 = \langle z^2\rangle$, $\langle
z^2\rangle$ is the second moment of the vertical component, and
$\alpha = \pi / \sqrt{12}$. With this normalization, the parameter $h$
is equivalent to the disk scale height $z_{\rm disk}$, $h \equiv
z_{\rm disk}$, when the second moment is calculated for all stars
sampled from the ${\rm sech}^2z/z_{\rm disk}$ model used for the
vertical density of our simulations (cf. Eq.~\eqref{eq:dens-disk}).

Once the average vertical action and vertical velocity dispersion are
computed for an ensemble of stars, Eq.~\eqref{eq:hdisk} is solved for
the thickness $h$ of that population of stars.

Fig.~\ref{fig:hdisk} displays the average value of the vertical
action $j_z$ as a function of the galactic radius at three different
times for both galaxy models (top panels). The solid lines show the
average vertical action calculated for a set of stars using the
algorithm described in Section~\ref{sec:preliminaries}. The dashed
lines illustrate the predicted vertical action using the epicycle
approximation for that set of stars. Clearly, the epicycle
approximation predicts values of the vertical action 15\% higher than
the estimates obtained with the numerical procedure. A similar trend
is observed in the numerical experiments described in
\citet{Solway2012}.

Then, Eq.~\eqref{eq:hdisk} is used to compute the thickness of the
disk $h$ at each radius, and the values obtained for the both galaxy
models are shown in Fig.~\ref{fig:hdisk} (bottom panels).  A similar
discrepancy is reflected in the estimate of the scale height at
different radii. The values derived in the epicycle approximation are
5\% larger than the values obtained by evaluating Eq.~\eqref{eq:action}
as previously described.

\begin{figure}
  \begin{center}
    \includegraphics[width=0.49\textwidth]{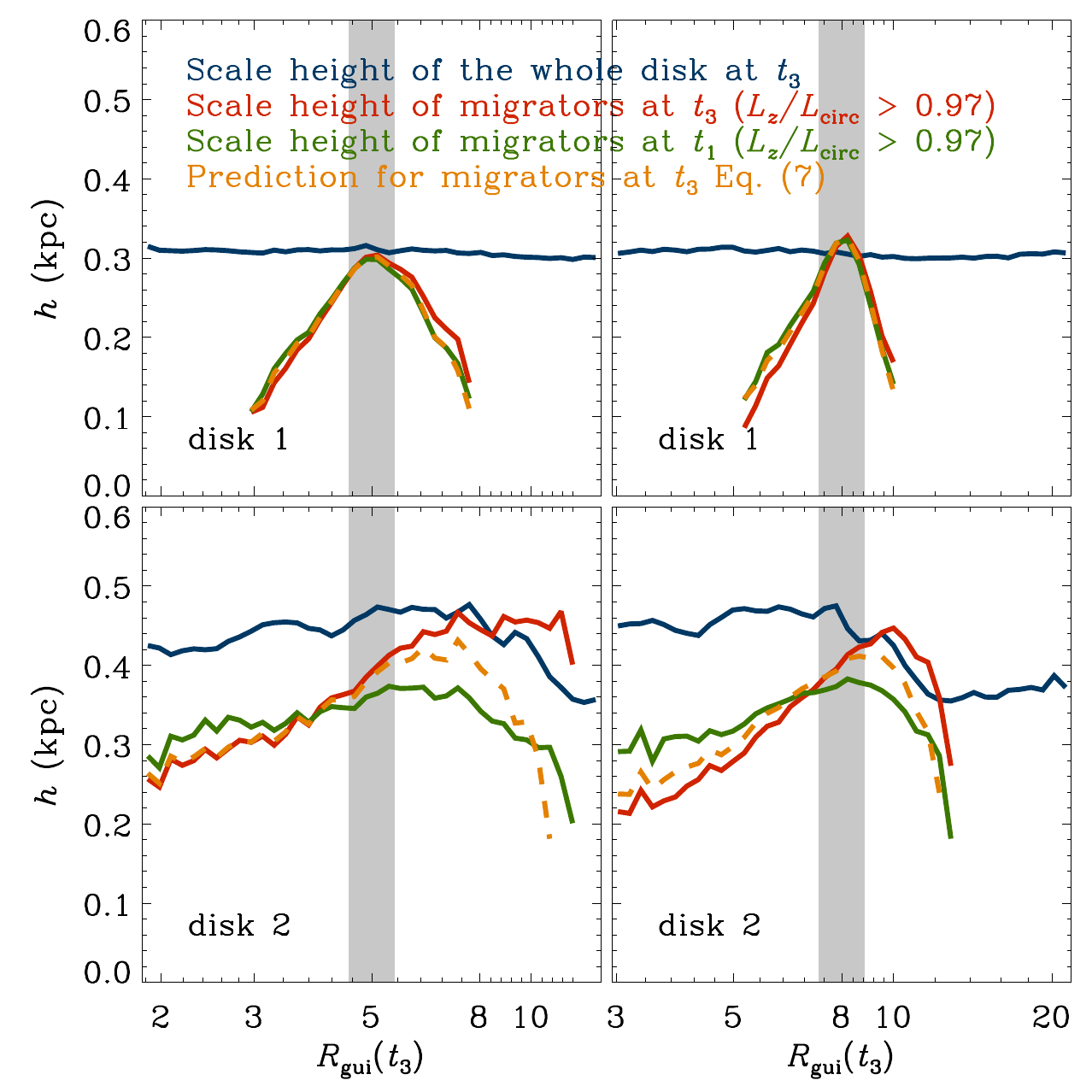}
  \end{center}
  \caption{Scale height $h$ profile for particles in circular orbits
    that migrate from $R_{\rm gui}(t_1) = 5.00 \pm 0.25$ kpc (left)
    and $R_{\rm gui}(t_1) = 8.00 \pm 0.40$ (right) as a function of
    their final guiding radii for the the multi-arm simulation (top)
    and the MW-like model (bottom). The scale height profile of
    the whole disk at $t_3$ is shown by the solid blue line. In green
    we show the thickness profile of the migrators at $t_1$ and in red
    the thickness at $t_3$. The orange dashed line is the result of
    using Eq.~\eqref{eq:hdisk} and assuming that the vertical action
    is conserved $\langle j_{z, \rm epi}(t_1) \rangle = \langle j_{z,
      \rm epi}(t_3) \rangle$.}
  \label{fig:hdisk-mig}
\end{figure}

In both simulations, after the disks evolved for 1 Gyr ($t_1$), we
selected a set of stars with guiding radii $R_{\rm gui} = 5.00 \pm
0.25$ kpc (left panels in Fig.~\ref{fig:hdisk-mig}) and $R_{\rm gui} =
8.00 \pm 0.40$ kpc (right panels), we follow these stars for two
billion years longer ($t_3$) and calculated their scale height $h$
profile as a function of their final position in the disk. This would
show if a subsample of migrating stars can change their thickness $h$
as they migrate.

We only select stars that are in nearly circular orbits $L_z/L_{\rm
  circ} > 0.97$, where $L_z$ is the vertical angular momentum and
$L_{\rm circ}$ is the angular momentum that a star has, should it move
in a circular orbit with the same energy. By adding this constraint at
both times $t_1$ and $t_3$ we ensure that the sample contains stars
that migrate while preserving their circular orbits
\citep{Sellwood2002}.

For this set of particles the average vertical action in the epicycle
approximation is also $j_z$ is computed at both times $t_1$ and
$t_3$. The thickness at both times is also included, in general stars
that move toward the center tend to reduce their scale height, whereas
stars that migrate outwards increase it. For comparison we also show
in blue the scale height of the disk as a function of radius,
calculated with the normalized RMS value of the $z$-coordinates of all
particles in the disk. This value matches the one reported as $z_{\rm
  disk}$ in table \ref{tab:params} (blue solid line).

The bias reported in \citep{VeraCiro2014} is observed for both
disks. This explains the fact that for stars that migrate outwards,
the slight increase in their vertical scale height is not enough to
create a thicker component at their final location. We also included
in Fig.~\ref{fig:hdisk-mig} the predicted thickness $h$ at final $t_3$
under the assumption that the action is conserved. It can clearly be
seen that for our \diskone{} \emph{no} thick distribution of stars is
either observed or predicted even when the average vertical action is
clearly conserved.

In the MW-like galaxy, \disktwo{}, a bar grows in the inner parts of
the disk, in qualitative agreement with simulations of
\citet{Brunetti2011}.  As a consequence of the bar formation and the
vigorous spiral activity there are not only significant changes of
angular momentum, but also a considerable in-plane heating that forces
stars in this simulation to deviate from a near-circular orbits.  There
are good reasons to believe that in this case the energy of vertical
motion is clearly not decoupled from that of the radial part of the
motion.  Indeed, the epicycle approximation is a poor description of
the motion of most particles, for which the radial and vertical
oscillations are neither harmonic nor are the energies of the two
oscillations decoupled.  Furthermore, the actions are constants of
motion under the conditions of slow changes of the potential. In the
multi-armed galaxy the low-amplitude arms represent tiny perturbations
of the potential, hence the vertical action might still be conserved,
although after several orbital periods of revolution the vertical
action is not conserved at the same level of accuracy.  However, in
the MW-like galaxy the high-amplitude spiral structure and the bar
formation seem to compromise the validity of these conditions.

\section{Summary and Conclusions}
\label{sec:conclusions}

We have presented quantitative estimates of the vertical action of
ensembles of stars in the presence of spiral activity.  The vertical
action of star particles has been evaluated in isolated disks of
galaxies with different spiral morphologies. The two galaxy models
consist of a low-mass multi-armed disk (\texttt{disk 1}) and a MW-like
disk that develops a bar after $\sim 3.5$ Gyr of evolution (\disktwo).

For the multi-armed disk galaxy, \diskone, we find that the vertical
action for an ensemble of stars is approximately a conserved quantity,
with a dispersion of fractional changes around its median of 20\%, in
agreement with the results reported in the literature. However, as
time progresses the action is conserved with less accuracy, as
indicated by the increase of the standard deviation up to the value of
50\% after 6 Gyr of evolution.  This also holds for a subset of
stars that radially migrates by resonant scattering at corotation.
These results have implications for the vertical structure of the disk
when stars radially migrate to the outer part of the disk.  In this
model stars migrating outwards are a heavily biased subset of stars
with preferentially low vertical velocity dispersion.  Thus, extreme
migrators outwards are characterized by having a have small amplitude
of their vertical excursion, independent of the conservation of the
vertical action.

In the MW-like galaxy model, \disktwo, the bar formation causes large
angular momentum changes, with significant in-plane and vertical
heating.  Thus, in the presence of the formation of the bar and
combined with significant disk heating, our results show that the
vertical action cannot be assumed a constant of motion.  Indeed, in
this case the fractional changes in the estimates of the action for a
set of stars progressively increase by time, with a standard deviation
that approaches 80\% after 6 Gyr of disk evolution.  If confirmed,
this result casts doubts on the possibility of predicting the vertical
structure of the disk for a set of extreme migrators outwards, by
assuming that their vertical action is conserved.

\vskip 3em

This is funded by NSF Grant No AST-1211258 and ATP NASA Grant No
NNX144AP53G.  ED gratefully acknowledges the support of the Alfred
P. Sloan Foundation.  We express our appreciation toward the Aspen
Center for Physics for their hospitality, funded by the NSF under
Grant No.  PHYS-1066293. Simulations have been run on the High
Performance Computing cluster provided by the Advanced Computing
Infrastructure (ACI) and Center for High Throughput Computing (CHTC)
at the University of Wisconsin. We thank Daniel Pfenniger, Jerry
Sellwood, Martin Weinberg, Victor Debattista, Rob Grand and Jonathan
Bird for the very helpful comments and constructive discussions.

\bibliographystyle{apj} 
\bibliography{refs}

\appendix

\section{Measuring the vertical action}
\label{sec:appendix}

The vertical action $j_z$ is defined as:

\begin{equation}\label{eq:action-apdx}
  j_z \equiv \frac{1}{2\pi}\int {\rm d}z\; {\rm d}v_z = \frac{1}{2\pi}\oint {\rm d}z\; v_z,
\end{equation}

\noindent where $z$ and $v_z$ describe the position and velocity along
the orbit, respectively.  If the potential is nearly axisymmetric this
quadrature can be calculated using the algorithm described in
\citep{Solway2012}. We adopted here a similar prescription as
explained below:

\begin{enumerate}

\item At a given time $t$ the gravitational potential $\Phi$, its
  radial gradient $\partial \Phi/\partial R$, and vertical gradient
  $\partial \Phi/\partial z$ are computed on a polar grid
  $(R,\theta,z)$ centered at the minimum of the (fixed) halo
  potential. The grid dimension is $N_R\times N_\theta \times N_z =
  128\times 128\times 128$, with the cylindrical coordinates ranging
  as follows: $\epsilon_\Phi < R < 30 {\;\rm kpc}$ (logarithmically
  spaced), $-\pi <\theta < \pi$ and $-6 \;{\rm kpc} < z < 6 \;{\rm
    kpc}$, where $\epsilon_\Phi = 50{\;\rm pc}$ is the softening
  length.  Note that these ranges span six times the scale-length and
  twenty times the scale height. The potential is numerically
  calculated at each node of the grid using a tree algorithm that
  includes up to the quadrupole contribution.  We do not impose a
  symmetry about the galactic mid-plane, in fact, modest asymmetries
  that can develop over time are naturally captured by the grid.

\item At each point $(R,z)$ the potential is calculated by averaging
  over the $N_\theta = 128$ azimuthal points:

  \begin{equation}
    \Phi(R,z) \equiv \frac{1}{N_\theta} \sum_{j=1}^{N_{\theta}} \Phi(R,\theta_j,z),
  \end{equation}
  
  \noindent Equivalent expressions are derived for $\partial
  \Phi/\partial R$ and $\partial \Phi/\partial z$. This approximation
  relies on the fact that the potential is nearly axisymmetric, this
  however may not be the case after if a bar grows at the center of
  the galaxy.
 
\end{enumerate}

Once the potential is computed at $t$ time, for each star particle the
following quantities are evaluated: the vertical angular momentum
$L_z$, the guiding radius $R_{\rm gui}$ as the solution to the
equation:

\begin{equation}
  L_z = R_{\rm gui} V_{\rm circ}(R_{\rm gui}),
\end{equation}

\noindent and the vertical action $j_z$, with the procedure described
below.

First, the position and velocity vectors are transformed to
cylindrical coordinates $(R,z)$. This point is integrated in the fixed
potential defined by the grid for a time $t_{\rm max} = 10^3 T_z$,
where $T_z = 2\pi /\nu$ is the period of vertical oscillations, and
$\nu$ is the vertical epicycle frequency \citep{Binney2008}

\begin{equation}
\nu^2 = \left. \frac{\partial^2 \Phi(R,z)}{\partial z^2} \right|_{z=0}.
\end{equation}

The vertical frequency can be easily calculated using five-point
stencil derivatives applied to the $\partial \Phi/\partial z$ grid.

The integration is performed using a symplectic integrator of order 4
\citep{Candy1991}. With this choice, typical deviations in the energy
are always smaller than $10^{-6}$ and show no secular growth over the
integration time $t_{\rm max}$. This procedure yields a set of
discrete points along the orbit of the form $\{t_i, R_i, z_i, v_{R,i},
v_{z,i}\}_i$.

During the integration of each star, every time the condition $R_i <
R_{\rm gui} < R_{i + 1}$, $v_{R,i} > 0$ is satisfied, the dynamical
states are changed from $\{ R(t), z(t), v_R(t), v_z(t)\} $ to $\{t(R),
z(R), v_R(R), v_z(R)\} $ which is numerically integrated in the range
$R_i \leq R \leq R_{\rm gui}$. With this change of coordinates, the
symplectic structure clearly breaks, therefore we have to use a
different (non-conservative) integrator. In this case we use the
explicit embedded Runge-Kutta Prince-Dormand (8, 9) method
\citep{Dormand1980} implemented in the
\texttt{\textsc{Gsl}}\footnote{\url{http://www.gnu.org/software/gsl/}}
library. After the \emph{exact} intersection with the $R = R_{\rm
  gui}$ plane is found, the integration is resumed from $R_{i + 1}$
with our symplectic method.

\begin{figure}
  \begin{center}
    \includegraphics[width=0.49\textwidth]{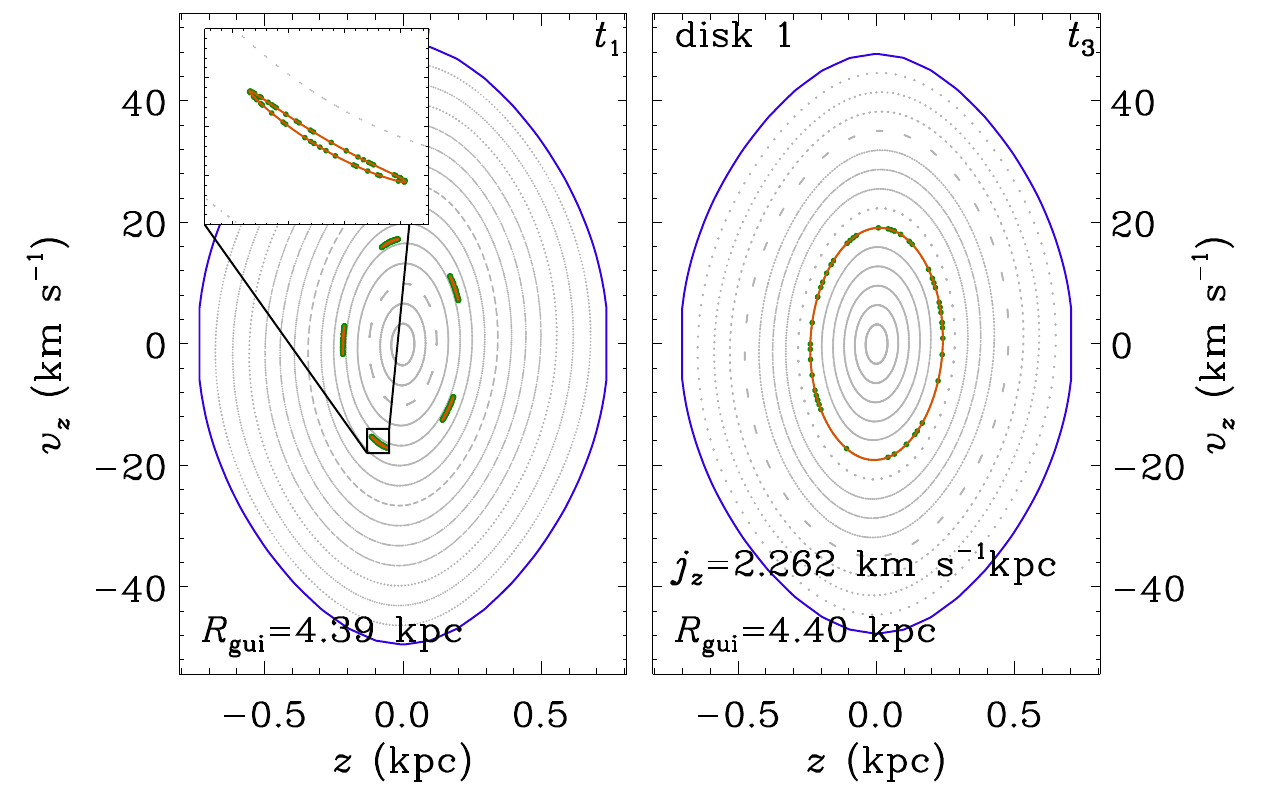}
  \end{center}
  \caption{Surface of section for a star particle measured at two different
    times. In the left panel the simulation has already developed
    spiral arms, the right panel is the surface of section after 2
    Gyr. In this specific example, $R_{\rm gui}(t_1) \approx R_{\rm gui}(t_3)$
    indicates that the particle has not migrated significantly. 
    However, the particle moves away from a resonant orbit, a
    situation that is easily identified with our scheme.}
  \label{fig:sos-apdx}
\end{figure}

In Fig.~\ref{fig:sos-apdx} we show an example for a random particle in our
\diskone{} simulation. The green points show the set obtained with the
procedure described above. The blue line represents the zero-velocity
curve, and gray points are other orbits, shown here to depict the
structure of the phase-space in our runs. At a first glance, the set
in the left panel resembles a simple closed structure in the
$(z,v_z)$-plane, however, a careful inspection reveals a more complex
structure. Indeed, the surface of section of this particle is composed
of five resonant islands. At later times, however, the orbit traces a
single closed structure (right panel). This transition is
non-adiabatic and as such we would like to devise a way to identify
these cases.

The task of calculating the vertical action for a given star is then
reduced to measuring the area enclosed by the set of points in the
$(z,v_z)$-plane found above as the star moves in the galaxy. To
achieve this, the points in the surface of section were connected with
the following procedure:

\begin{enumerate}
\item In the $(z,v_z)$-plane a random point $p_0 \equiv (z_0,v_{z,0})$
  is selected and its closest neighbor $p_1$ is found. For simplicity,
  both coordinates $(z,v_z)$ are rescaled in the interval $(-1,1)$.

\item The closest $N$ points to $p_1$, $\{q_k\}_{k=1}^N$ that have not
  been already included in a previous set are then selected. For each
  point $q_k$ the angles between the segments $\overline{p_0p_1}$ and
  $\overline{p_1q_k}$ are defined.

\item The point with minimum angle weighted by the distance is found
  and defined as the the new $p_1$. We noted that in our
  implementation the weights $\sim|\overline{p_1q_k}|^3$ perform
  reasonably well.

\item Steps 2 and 3 are repeated until $p_0$ is the next point to
  be included in the cycle.

\item Step 1 is then repeated until all points have been included in
  our analysis.

\end{enumerate}

This algorithm then yields a set of unconnected loops in the surface
of section defined by $R=R_{\rm gui}$ and $v_R>0$. In the cases where
only one loop is present the action is just the area of such cycle and
can be calculated using the popular formula based on the Green
theorem. Otherwise, the particle is trapped in a resonance and it is
no longer considered in our results. This implementation uses $N=16$
and rejects any cycle that has fewer points than this threshold. In
addition, orbits for which the distance between the last and first
point is larger than five times the average distance between the other
points in the cycle are rejected. This choice avoids cases in which
the algorithm ends before closing a loop, a circumstance that occurs
for stars whose orbit takes them close to the zero-velocity curve. If
the orbit is composed of several loops (e.g. left panel in
Fig.~\ref{fig:sos}) and the procedure fails to calculate the area for
one of those loops, then the orbit is rejected altogether.

Fig.~\ref{fig:sos-apdx} shows the results of the implementation of our
algorithm to the particles of our simulations, as illustrated by the
red line of the inset of the left panel.  In our numerical experiments
the procedure successfully computes the vertical action for $\sim 85\%$
of the particles located at a distance of $R\gtrsim 2$ kpc from the
disk center, and $\sim 60\%$ of particles orbiting at smaller radii.

\end{document}